\documentclass[a4paper,fleqn]{cas-dc}

\usepackage[numbers]{natbib}

\def\tsc#1{\csdef{#1}{\textsc{\lowercase{#1}}\xspace}}
\tsc{WGM}
\tsc{QE}
\tsc{EP}
\tsc{kMS}
\tsc{BEC}
\tsc{DE}

\begin{document}
\let\WriteBookmarks\relax
\def\floatpagepagefraction{1}
\def\textpagefraction{.001}
\shorttitle{}
\shortauthors{}

\title [mode = title]{Adiabatic limit of RKKY range function in one dimension} 
\tnotemark[1]

\tnotetext[1]{This document is the results of the research project funded by Universitas Indonesia through PUTI Q2 Grant No. NKB-1663/UN2.RST/ HKP.05.00/2020.}

\author[1]{Adam B. Cahaya}[
orcid=0000-0002-2068-9613]
\cormark[1]
\fnmark[1]
\ead{adam@sci.ui.ac.id}

\credit{Conceptualization, Methodology, Writing - Review & Editing, Funding acquisition }

\address[1]{Department of Physics, Faculty of Mathematics and Natural Sciences, Universitas Indonesia, Depok 16424, Indonesia, (62)21-7872610}

\begin{abstract}
The RKKY interaction is an important theoretical model for indirect exchange interaction in magnetic multilayer. The expression for RKKY range function in three dimension and lower has been derived in the 1950s. However, the expression for one dimension is still studied in recent years, due to its strong singularity. By using an adiabatic limit of retarded Green's function form that directly related to RKKY interaction in one dimension, we decompose the singularity and recover the range function. Furthermore, we show in adiabatic limit, RKKY interaction also induces one-dimensional spin pumping and distance-dependent magnetic damping.

\end{abstract}




	\maketitle


\section{Introduction}

The discovery of giant magnetoresistance effect has led to application on  magnetic memory. It improves the speed of reading process of magnetic memory\cite{reviewGMR}. Since then, the research area of spintronics that analyzes spin and charge dynamics has emerged \cite{SpinReview,Spintronics}. 
To further improve the magnetic memory, one of the aims of spintronics research is an effective manipulation of magnetization \cite{SpinRAM,STTRAM}. By utilizing the spin-degree of freedom, magnetization can be manipulated by various methods, such as spin current \cite{PhysRevB.72.024426}, electric current \cite{SLONCZEWSKI,spinHall} and voltage \cite{AlePRL,Maruyama2009,Nozaki2012,LeonPRB}. 

In magnetic multilayer, the indirect exchange is a widely-studied method to control magnetization. In magnetic multilayer, it has been observed that there is an interlayer exchange interaction. The interlayer exchange interaction is an indirect interaction between ferromagnetic layers that is mediated by conducting electron of the sandwiched non-magnetic layer~\cite{RKKYBruno}. The indirect interaction is called RKKY (Ruderman - Kittel - Kasuya - 	Yosida) interaction which is named after the discoverers \cite{RK,K,Y}.
\begin{align}
H_{RKKY}=-J_{RKKY} \left(\left| x_1- x_2\right|\right) \textbf{S}_1\cdot \textbf{S}_2,
\end{align}
where $J_{RKKY} \left(\left| x_1- x_2\right|\right)$ is an exchange constant that depends on the distance between the spins $\textbf{S}_1$ at position $ x_1$ and $ \textbf{S}_2$ at position $ x_2$. 
The exchange constant of RKKY interaction in three-dimensional system depends on the distance between the two spins
\begin{align}
J^\mathrm{3D}_\mathrm{RKKY}(x)\propto \frac{\sin 2k_Fx}{2k_Fx^4} - \frac{\cos 2k_Fx}{x^3},
\end{align}
where $k_F\propto\sqrt{\epsilon_F}$ is the Fermi wave-vector and $\epsilon_F$ is the Fermi energy of the conduction electron. The right side of the above equation is often called RKKY range function. The trigonometric functions indicate a spatial-oscillation of $J^{3d}_\mathrm{RKKY}$. The coupling is ferromagnetic (antiferromagnetic) when the sign of $J_{RKKY}$ is positive (negative). 

The expression for lower dimensions has been showed to have the following forms \cite{Litvinov,Yafet}.
\begin{align}
J^\mathrm{2D}_\mathrm{RKKY}(x)\propto&  J_0(k_Fx)Y_0(k_Fx)+J_1(k_Fx)Y_1(k_Fx),\notag\\
J^\mathrm{1D}_\mathrm{RKKY}(x)\propto& \frac{\pi}{2} -\mathrm{Si}(2k_Fx).
\end{align} 
Here $\mathrm{Si}(x)$ is the Sine integral \cite{SineI}. $J_n(x)$ and $Y_n(x)$ is the $n$-th Bessel function of the first kind \cite{Bessel1} and the second kind \cite{Bessel2}, respectively. 
The original theoretical description by Kittel shows that $J_{RKKY}(x)$ is proportional to the static magnetic susceptibility $\chi\left(x\right)$\cite{Kittel}. 
The expression of $\chi\left(x\right)$ involves double integration in the momentum space $k$ and $q$. This integral has singularities at $| k+ q|^2=q^2$. For one dimensional  RKKY interaction, the methods for derivation of $\chi(x)$ are still studied in recent years, due to a strong singularity at $k=q=0$ \cite{RusinJMMM,RusinPRB,PhysRevB.72.033411}.
In a more realistic case, the spin impurities are not static. They can have a finite precession. In this case, $\chi$ also depends on the time. 

Dynamic correction for $\chi$ has been shown to associate with spin pumping in three-dimensional magnetic multilayer \cite{SimanekPRB,CahayaCF}. While 
spin pumping in two systems has also been theoretically studied \cite{Rahimi2015,PhysRevB.94.205428}, spin pumping in one system has not been well-described yet. In this article, we predict the mechanism of spin pumping in one dimensional system by studying the dynamic correction of $\chi$. 

\begin{figure}[b]
\centering
\includegraphics[width=\columnwidth]{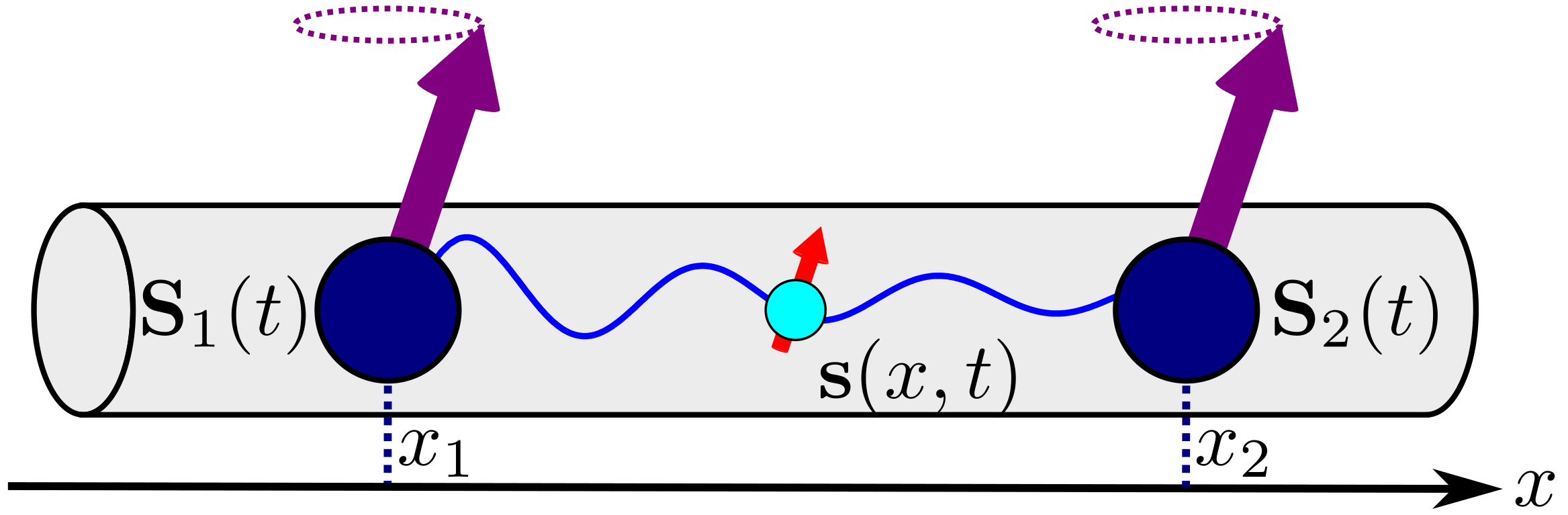}
\caption{Indirect interaction between two magnetic impurities $\textbf{S}_1$ and $\textbf{S}_2$ in one dimensional system mediated by conduction spin $\textbf{s}$. 
\label{figWire}}
\end{figure}

This article is organized as follows. In Sec.~\ref{Sec:dynamicChi}, we validate the expression for $\chi(x,t)$. In Sec.~\ref{Sec:singular}, we discuss the singularities that appear in the expression of $\chi(x,t)$ and discuss its adiabatic limit. In Sec.~\ref{Sec.damping} we show that the adiabatic limit of $\chi(x,t)$ can describe spin current pumping in one dimensional system. Lastly, we summarize our result in Sec.~\ref{Sec:conclusion}.

\section{Dynamic magnetic susceptibility}
\label{Sec:dynamicChi}
In RKKY interaction, the indirect exchange interaction between spins $\textbf{S}_1$ at position $x_1$ and $ \textbf{S}_2$ at position $x_2$ is mediated by the spin of conduction electron $\textbf{s}(x)$ (see Fig.~\ref{figWire}). 
According to Kubo formula, $\textbf{s}(x,t)$ can be found from the linear order in perturbation \cite{Kubo1,Kubo2,Doniach}
\begin{equation}
\textbf{s}(x,t)= -\frac{i}{\hbar}\int_{-\infty}^t dt'\left<\left[\textbf{s}(x,t),H(t')\right]\right>, \label{Eq.Kubo}
\end{equation}
where $H(t)$ is  the exchange interaction between the impurity and conduction spins, which can be written in the following $s-d$ Hamiltonian \cite{Kondo,Larsen1,Larsen2}
\begin{align}
H(t)=-J\sum_{n=1,2}\textbf{S}_n(t)\cdot \textbf{s}(x_n,t). \label{Eq.Hamiltonian}
\end{align}
Here $J$ is a direct exchange constant between the impurity spin and conduction spin. 
$\textbf{s}(x_i,t)$ can be obtained by studying the retarded response of conduction spin to the spin impurities. Similar approached has been used for adiabatic correction in three dimensional system \cite{SimanekPRB,CahayaCF}.

By substituting Eq.~\ref{Eq.Hamiltonian} into Eq.~\ref{Eq.Kubo}, $\textbf{s}(x,t)$ can be expressed in terms of the susceptibility function $\chi(x - x',t-t')$ and  magnetic field from impurity spins $\textbf{B}(x',t')=J\sum_n\textbf{S}_n(t')\delta(x'-x_n)$.
\begin{eqnarray}
&s_i(x,t)= \int dx' \int dt' \chi_{ij}(x - x',t-t') B_j(x',t')\notag\\
&= J\sum_n\int dx'\int dt' \chi_{ij}(x-x',t-t') S_{nj}(t')\delta(x'-x_n). \label{Eq.sigma}
\end{eqnarray} 
The above convolution expression of $s_i$ can be simplified by expressing it in Fourier space $(k,\omega)$
\begin{align}
s_i(k,\omega)= J\sum_n e^{ik(x-x_n)}\chi_{ij}(k,\omega) {S}_{nj}(\omega).
\end{align}
In linear response theory, the susceptibility is defined as the following retarded response function \cite{PhysRevLett.17.750,CahayaSTT}
\begin{equation}
\chi_{ij}(x-x',t-t')=i\Theta(t-t') \left<\left[s_i(x,t),s_j(x',t')\right]\right>.
\end{equation}
The retarded response is also used in Refs.~\cite{CahayaCF,PhysRevLett.17.750}. Here $\Theta(t)$ is Heaviside step function that can be written by its Fourier transform $\Theta(\omega)$ \cite{Bracewell1921-2007}
\begin{equation}
\Theta(t) =  \int_{-\infty}^\infty \frac{d\omega }{2\pi} e^{-i\omega t} \Theta(\omega)
= \lim _{\eta\to 0^+} \int_{-\infty}^\infty \frac{d\omega }{2\pi}   \frac{e^{-i\omega t}}{\eta-i\omega}.
\end{equation}
Its derivative is
\begin{equation}
\Theta'(t) =
\lim _{\eta\to 0^+} \int_{-\infty}^\infty \frac{d\omega }{2\pi}   \frac{-i\omega e^{-i\omega t}}{\eta-i\omega}
=\delta(t)-\lim _{\eta\to 0^+} \eta\Theta(t). \label{Eq.DiffStep}
\end{equation}
$s_i(x,t)$ in second quantization can be written as follows \cite{Doniach}.
\begin{equation}
s_i(x,t)=\sum_{\alpha\beta} \iint \frac{dkdq}{(2\pi)^2}e^{iqx} a^\dagger_{k+q,\alpha}(t)\left(\sigma_i\right)_{\alpha\beta} a_{k,\beta}(t), \label{Eq.skt}
\end{equation}
$a_{k\alpha}$ and $a^\dagger_{k\alpha}$ are the annihilation and creation operators of electron with wavevector $k$ and spin $\alpha$ in second quantization, respectively. 
Here $\sigma_i$ ($i=x,y,z$) are the Pauli matrices.
\begin{align}
\boldsymbol{\sigma}=\left(\left[\begin{array}{cc}
0& 1\\ 1& 0
\end{array}\right],
\left[\begin{array}{cc}
0& -i\\ i& 0
\end{array}\right],
\left[\begin{array}{cc}
1& 0\\ 0& -1
\end{array}\right]\right).
\end{align}
For convenience, we can write $\chi_{ij}(x,t)$ in terms of $\Gamma_{ij}(k,q,t)$ as follows.
\begin{align}
\chi_{ij}(x,t) =& \iint \frac{dkdq}{(2\pi)^2} e^{iqx} \Gamma_{ij} (k,q,t), \notag\\
\Gamma_{ij} (k,q,t)= & i\Theta(t) \sum_{\alpha\beta} \left(\sigma_{i}\right)_{\alpha\beta} \left<\left[a_{k+q,\alpha}(t)a_{k,\beta}(t),s_j(0,0)\right]\right>.
\end{align} 

We can evaluate the time derivative of $\Gamma_{ij}$ by using Eq.~\ref{Eq.DiffStep}.
\begin{align}
&\frac{\partial\Gamma_{ij}(k,q,t)}{\partial t}= \lim_{\eta\to 0} \eta \Gamma_{ij}(k,q,t)\notag\\
&+ i\delta(t) \sum_{\alpha\beta} \left(\sigma_{i}\right)_{\alpha\beta} \left<\left[a_{k+q,\alpha}(t)a_{k,\beta}(t),s_j(0,0)\right]\right>\notag\\
&+i\Theta(t) \sum_{\alpha\beta} \left(\sigma_{i}\right)_{\alpha\beta} \frac{\partial}{\partial t}\left<\left[ a_{k+q,\alpha}(t)a_{k,\beta}(t),s_j(0,0)\right]\right> . \label{Eq.diffGamma}
\end{align}
By using Eq.~\ref{Eq.skt}, relation of Pauli matrices $\sigma_a\sigma_b=\delta_{ab}I+i\varepsilon_{abc}\sigma_c$ and commutation relation of the annihilation and creation operator, one can evaluate the second term in the right side \cite{Doniach}
\begin{align}
&\sum_{\alpha\beta}\left(\sigma_i\right)_{\alpha\beta} \left<\left[a^\dagger_{k+q,\alpha}(t)a_{k,\beta}(t),s_j(0,0)\right]\right>
\notag\\
&\simeq\sum_{\alpha\beta}\left(\sigma_i\right)_{\alpha\beta} \left<\left[a^\dagger_{k+q,\alpha}(0)a_{k,\beta}(0),s_j(0,0)\right]\right>
\notag\\
&=\sum_{\alpha\beta} \left(f_{k+q,\alpha}-f_{k,\beta}\right)\delta_{ij}\delta_{\alpha\beta}. \label{Eq.14}
\end{align}
where $f_{k,\alpha}$ is the Fermi-Dirac distribution of electron with energy $\epsilon_k$ and spin $\alpha$. 
The time derivative of the last term in Eq.~\ref{Eq.diffGamma} can be evaluated as follows
\begin{equation}
\frac{\partial\left( a_{k+q,\alpha}(t)a_{k,\beta}(t)\right)}{\partial t}
= -i\left[a^\dagger_{p+k,\alpha}(t)a_{p,\beta}(t), H_0\right],
\end{equation}
Here $H_0$ is the unperturbed Hamiltonian of free electron system.  
\begin{align}
H_0= \sum_{\alpha}\int \frac{dk}{2\pi} \epsilon_k a^\dagger_{k,\alpha}(t) a_{k,\alpha}(t).
\end{align}
For simplicity, we set $\epsilon_k=k^2$. By using the commutation relation of the creation and annihilation operators, one can show the following relations \cite{Doniach}
\begin{align}
&\left[a^\dagger_{k+q,\alpha} a_{k,\beta},H_0\right] =  \left(\epsilon_{k}-\epsilon_{k+q} \right)a^\dagger_{k+q,\alpha} a_{k,\beta}. \label{Eq.17}
\end{align}
Eq.~\ref{Eq.diffGamma} can now be simplified by using Eqs.~\ref{Eq.14} and \ref{Eq.17} 
\begin{align}
&\left(\frac{\partial}{\partial t}- \lim_{\eta\to 0} \eta +i\left(\epsilon_{k}  - \epsilon_{k+q}\right) \right)\Gamma_{ij}(k,q,t)\notag\\
&= i\delta(t) \left(f_{k+q}-f_{k}\right)\delta_{ij}, \label{Eq.linierGamma}
\end{align}
where $f_k=f_{k,\uparrow}+f_{k,\downarrow}$.  
By taking its Fourier transform, we can linearize Eq.~\ref{Eq.linierGamma} and obtain $\Gamma_{ij}$
\begin{align}
\Gamma_{ij}(k,q,\omega)=& \delta_{ij}\lim_{\eta\to 0}  \frac{f_k-f_{k+ q}}{\epsilon_{k+ q}-\epsilon_k+\omega+i\eta}.
\end{align}
We can then arrive at the same expression as in Refs.~\cite{CahayaCF,Sitorus_2021}
\begin{align}
\chi( x,t)=&\int \frac{d\omega}{2\pi} e^{-i(\omega +i\eta)t}\chi(x,\omega),\label{Eq.chixtw}\\
\chi( x,\omega)=&\lim_{\eta\to 0}\int \frac{dqdk}{(2\pi)^2}e^{i q\cdot x} \frac{f_k-f_{k+ q}}{\epsilon_{k+ q}-\epsilon_k+\omega+i\eta}. 
\label{Eq.chiAdiabatic}
\end{align}
The inclusion of $\eta\to +0$ 
creates an imaginary term according to the following Sokhotski–Plemelj formula \cite{Carcione2019}
\begin{equation}
\lim_{\eta\to 0}\int dz\frac{f(z)}{z-z_0+i\eta}= \int dz\frac{f(z)}{z-z_0} + i\pi f(z_0) \label{Eq.PS}.
\end{equation}
See Appendix~\ref{Sec:appendix-spatial-ImSusc} for derivation of imaginary part of $\chi(q,\omega)$ using Sokhotski–Plemelj formula.

Let us focus on the frequency dependent magnetic susceptibility $\chi(x,\omega)$. 
By substituting $k+ q=k'$ in $f_{k+ q}$, we can arrive at the following expression.
\begin{align}
\chi(x,\omega)=&\int \frac{dq}{2\pi}e^{i q\cdot x}\int \frac{dk}{2\pi} \frac{f_k}{(k+ q)^2-k^2+\omega+i\eta}\notag\\
&-\int \frac{dq}{2\pi}e^{i q\cdot x}\int \frac{dk'}{2\pi} \frac{f_{k'}}{k'^2-(k'-q)^2+\omega+i\eta}\notag\\
=&\frac{1}{2\pi^2}\int_{-k_F}^{k_F} dk \int_{-\infty}^{\infty} dq e^{iqx}\left( \frac{1}{(q-q_1)(q-q_2)}\right.\notag\\
&+\left.\frac{1}{(q-q_3)(q-q_4)}\right), \label{Eq.chirw}
\end{align}
where the poles $q_{1,2,3,4}$ are
\begin{align}
q_{1,2}=& k\pm \sqrt{k^2-\omega-i\eta},\notag\\
q_{3,4}=& k\pm \sqrt{k^2+\omega+i\eta}.
\end{align}
When $\omega$ and $\eta$ approach zero, the poles converge to $q=0$ and $q=2k$, which are the poles of the static susceptibility in Ref.~\cite{Kittel}.

\section{Singularities in complex plane }
\label{Sec:singular}

\begin{figure}[t]
\centering
\includegraphics[width=\columnwidth]{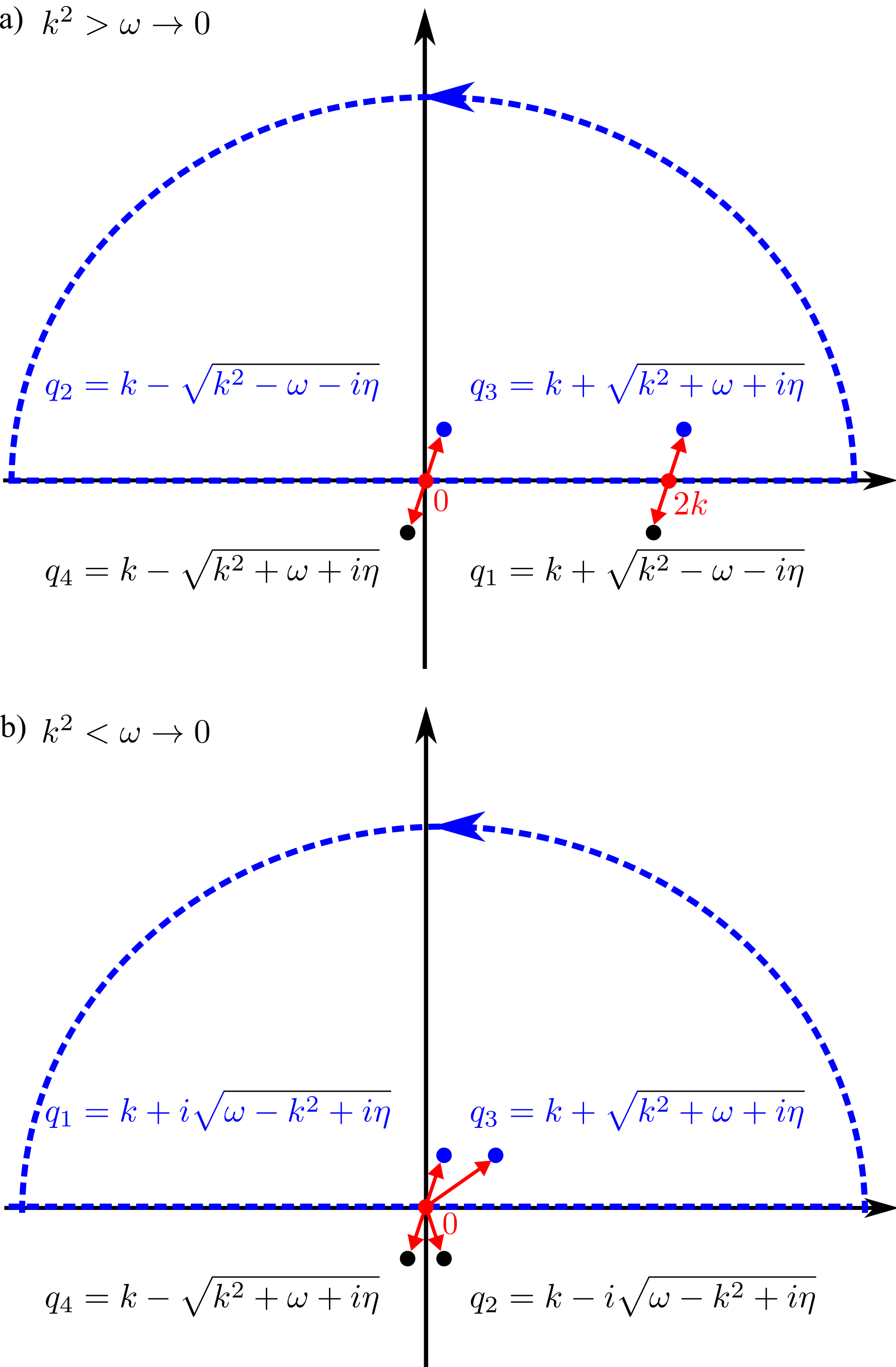}
\caption{Poles $q_1,q_2,q_3$ and $q_4$ at the complex plane. Blue contours are used to evaluate the integral in Eq.~\ref{Eq.contour}. 
Red dots are the poles for static susceptibility ($\omega=\eta=0$) used in Ref.~\cite{Kittel}. (a) When $k^2>\omega$, $q_2$ and $q_3$ have positive imaginary part. In the limit of $\omega=\eta=0$, the poles converge to $q=0$ or $q=2k$. (b) When $k^2<\omega$, $q_1$ and $q_3$ have positive imaginary part.  In this case, when $\omega=\eta=0$, all the poles converge to a single singularity at $q=k=0$. 
\label{figContour}}
\end{figure}

Fig.~\ref{figContour} illustrates that by using finite $\omega$ and $\eta$ the poles of the static susceptibility ($q=0$ and $q=2k$) can be decomposed into $q_{1,2,3,4}$. In particular, Fig.~\ref{figContour}b illustrates the decomposition of the strong singularity at $q=k=0$ \cite{RusinJMMM,RusinPRB} into 4 simple poles $q_{1,2,3,4}$.

Integral over $q$ in Eq.~\ref{Eq.chirw} can be evaluated by using Cauchy integral. 
For positive $x$, poles in the upper with positive imaginary part contribute to the value of the integral \cite{Bak2010}. 
\begin{align}
\int_{-\infty}^\infty dz\ e^{ixz}f(z)=\oint_C dz\ e^{ixz}f(z).\label{Eq.contour}
\end{align}
Here, closed path integral is taken over the contour in Fig.~\ref{figContour}. Depending on the relative value of $k^2$ and $\omega$, the sign of the imaginary part of $q_1$ and $q_2$ changes sign, as illustrated in Fig.~\ref{figContour}. 
Therefore, $\chi(x,\omega)$ in $\eta\to 0$ limit can be separated into four terms as follows.
\begin{align}
\chi(x,\omega)&= \int^{-\sqrt{w}}_{-k_F} \frac{dk}{\pi} i \frac{e^{iq_2x}}{q_2-q_1} +\int_{-\sqrt{w}}^{\sqrt{w}} \frac{dk}{\pi} i \frac{e^{iq_1x}}{q_1-q_2} 
\notag\\
&+\int_{\sqrt{w}}^{k_F} \frac{dk}{\pi} i \frac{e^{iq_2x}}{q_2-q_1} 
+\int_{-k_F}^{k_F} \frac{dk}{\pi} i \frac{e^{iq_3x}}{q_3-q_4}\notag\\
\equiv& \chi_1(x,\omega)+\chi_2(x,\omega)+\chi_3(x,\omega)+\chi_4 (x,\omega). \label{Eq.chitot}
\end{align}

Here, $\chi_{1,2,3,4}(r,\omega)$ can be evaluated separately.
By substituting $q_1$ and $q_2$, we can obtain $\chi_1(x,\omega)$ 
\begin{align}
\chi_1(x,\omega)&=\int^{-\sqrt{w}}_{-k_F} \frac{dk}{2\pi} i \frac{e^{ix\left(k-\sqrt{k^2-\omega}\right)}}{-\sqrt{k^2-\omega}} \notag\\
&=-i\int_{x\sqrt{\omega}}^{x\left(k_F+\sqrt{k_F^2-\omega}\right)} \frac{dz}{2\pi z} e^{-iz} ,\label{Eq.interval1}
\end{align}
and $\chi_2(x,\omega)$
\begin{align}
\chi_2(x,\omega)&=\int_{-\sqrt{w}}^{\sqrt{w}} \frac{dk}{2\pi} i  \frac{e^{ix\left(k+i\sqrt{\omega-k^2}\right)}}{i\sqrt{\omega-k^2}}\notag\\ 
&=\int_{-\pi/2}^{\pi/2} \frac{d\theta e^{ix\sqrt{\omega}e^{-i\theta}}}{2\pi} .\label{Eq.interval2}
\end{align}
By substituting $q_3$ and $q_4$, we can obtain $\chi_3(x,\omega)$ 
\begin{align}
\chi_3(x,\omega)&=\int_{\sqrt{w}}^{k_F} \frac{dk}{2\pi} i  \frac{e^{ix\left(k-\sqrt{k^2-\omega}\right)}}{-\sqrt{k^2-\omega}}\notag \\
&=i\int_{x\sqrt{\omega}}^{x\left(k_F-\sqrt{k_F^2-\omega}\right)} \frac{dz}{2\pi z} e^{iz} \label{Eq.interval3},
\end{align}
and $\chi_4(x,\omega)$ 
\begin{align}
\chi_4(x,\omega)&=\int_{-k_F}^{k_F} \frac{dk}{2\pi} i  \frac{e^{ix\left(k+\sqrt{k^2+\omega}\right)}}{\sqrt{k^2+\omega}}\notag \\
&=i\int_{x\left(-k_F+\sqrt{k_F^2+\omega}\right)}^{x\left(k_F+\sqrt{k_F^2+\omega}\right)} \frac{dz}{2\pi z} e^{iz}. \label{Eq.interval4}
\end{align}
In the adiabatic limit $\omega\ll k_F^2$, the integration interval of Eqs.~\ref{Eq.interval1}-\ref{Eq.interval4} can be expanded into its leading terms, as summarized in Table~\ref{Table.interval}. Then, we can use the following approximation
\begin{equation}
\lim_{x_{1,2}\to 0}\int_{a+x_1}^{b+x_2}dx f(x)=x_2 f(b) - x_1 f(a)+\int_{a}^{b}dxf(x) , \label{Eq.approximation}
\end{equation}
to determine the expression of $\chi_{1,2,3,4}(x,\omega)$ to the first order of $\omega$.

\begin{align}
\chi_1(x,\omega)=&i\omega \frac{e^{-i2k_Fx}}{8\pi k_F^2} +\int_{0}^{2k_Fx} \frac{dz}{\pi z} \frac{e^{-iz}}{2i} ,\notag\\
\chi_2(x,\omega)=& \frac{1}{2} ,\notag\\
\chi_3(x,\omega)=& 
\frac{i\omega}{8\pi k_F^2} -\int_{0}^{\frac{\omega x}{2k_F}} \frac{dz}{\pi z} \frac{e^{iz}}{2i},\notag\\
\chi_4(x,\omega)=& 
\frac{i\omega}{8\pi k_F^2} + i\omega \frac{e^{i2k_Fx}}{8\pi k_F^2} -\int_{\frac{\omega x}{2k_F}}^{2k_Fx} \frac{dz}{\pi z} \frac{e^{iz}}{2i} \label{Eq.chi1-4},
\end{align}

\begin{figure}[h]
\centering
\includegraphics[width=\columnwidth]{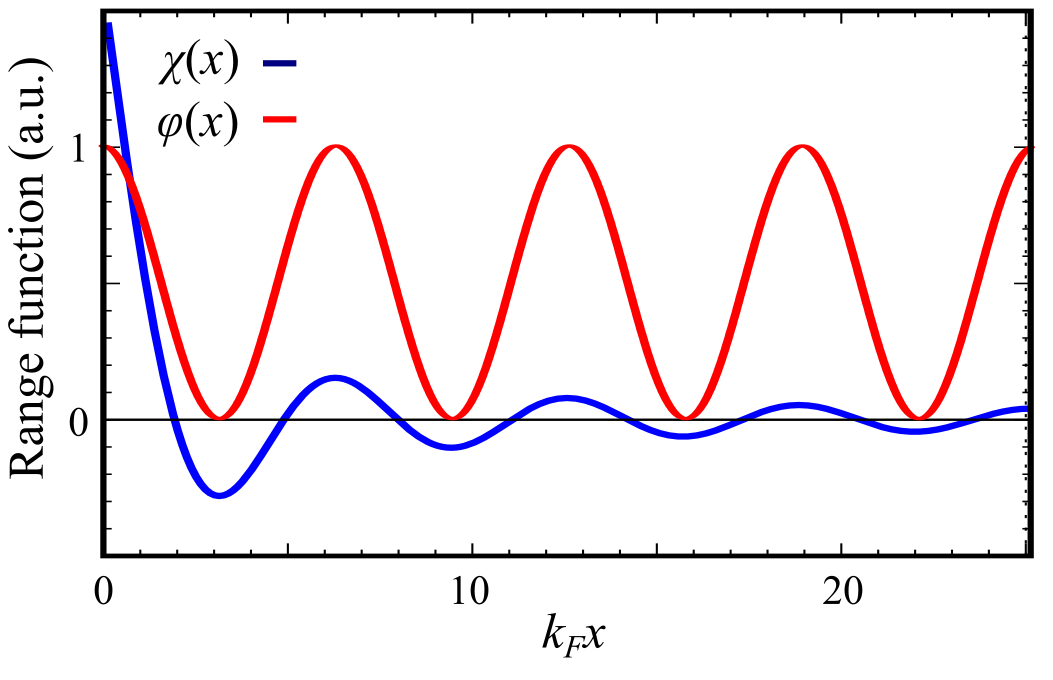}
\caption{Spatial oscillation of real and imaginary parts of frequency dependent susceptibility $\chi(x,w)$. In adiabatic limit $\lim_{\omega\ll k_F^2} \chi(x,w)= \chi(x) + i\omega \varphi(x)$. Blue line is  $\chi(x)\propto \pi/2-\mathrm{Si}(2k_Fx)$, which is the range function of the RKKY interaction in one dimension. On the other hand, red line is $\varphi(x)\propto\cos^2k_Fx$, the adiabatic correction that has spatial oscillation with similar period to the static term but does not change sign.
\label{figReIm}}
\end{figure}

By substituting Eq.~\ref{Eq.chi1-4} back to Eq.~\ref{Eq.chitot}, we can obtain complex-valued $\chi(x,\omega)$
\begin{align}
\chi(x,\omega)=& \chi(x)+i\omega \varphi(x) , \label{Eq.chiphi}
\end{align}
where 
\begin{align}
\chi(x)=& \frac{1}{2}-\int_{0}^{2k_Fx} \frac{dz}{\pi z} \frac{e^{iz}-e^{-iz}}{2i} \notag\\
=&\frac{1}{\pi}\left(\frac{\pi}{2}-\mathrm{Si} (2k_Fx)\right), 
\end{align}
is the static  magnetic susceptibility that gives the range function of the RKKY interaction in one dimension and 
\begin{align}
\varphi(x)=& \frac{1+\cos 2k_Fx}{4\pi k_F^2} = \frac{\cos^2 k_Fx}{2\pi k_F^2} . \label{Eq.varphi}
\end{align}
is the adiabatic correction of the susceptibility. $\varphi(x)$ has spatial oscillation with period similar to the static term.  However, there is no sign change, as illustrated in Fig.~\ref{figReIm}. In the next section, we show that this term induces distance dependency of the damping of RKKY coupled spins.

\begin{table}[h]
\caption{Interval of integration for evaluation of $\chi_{j}(x,\omega)$, $j=1,2,3,4$ in adiabatic limit $\omega\ll k_F^2$ for approximation of $\chi(x,\omega)$ using Eq.~\ref{Eq.approximation} \label{Table.interval}}
\begin{tabular}{cc}
\hline\\[-2ex]
$\chi_j(x,\omega)$ & integral interval \\
\hline\\[-2ex]
$\chi_1(x,\omega)$ & $\left[{x\sqrt{\omega}},{2k_Fx-\frac{\omega x}{2k_F}}\right]$ \\[1ex]
$\chi_2(x,\omega)$ & $\left[-\frac{\pi}{2},\frac{\pi}{2}\right]$ \\
$\chi_3(x,\omega)$ & $\left[{x\sqrt{\omega}},{\frac{\omega x}{2k_F}+\frac{\omega^2x}{8k_F^3}}\right]$ \\
$\chi_4(x,\omega)$ & $\left[{\frac{\omega x}{2k_F}-\frac{\omega^2x}{8k_F^3}},{2k_Fx+\frac{\omega x}{2k_F}}\right]$ \\
\hline
\end{tabular}
\end{table}

\section{Phenomenons related to the dissipative susceptibility $\varphi$}
\label{Sec.damping}

The expression of $s(x,t)$ can be obtained by rewriting Eq.~\ref{Eq.sigma} in terms of integration in the frequency domain using Fourier transform as follows
\begin{align}
s(x,t)=& J\sum_i\int d\omega e^{-i\omega t}\chi(x-x_i,\omega) \textbf{S}_i(\omega)\notag\\
=& J\sum_i\chi(x-x_i) \textbf{S}_i(t) - J\sum_i\varphi(x-x_i) \dot{\textbf{S}}_i(t) \label{Eq.sigmaReIm}.
\end{align}
Here, we can see that $\varphi$ gives an additional term that depends on the time derivative of the impurity spin $\dot{\textbf{S}}$. 
In this section we demonstrate novel phenomena that emerges due to a new term in RKKY susceptibility $\varphi$:
\begin{enumerate}
\item Spin pumping by a single impurity; and
\item Indirect damping between two impurities. 
\end{enumerate}

\subsection{Spin pumping by a single magnetic impurity}
\label{Sec.spinpumping}

Spin pumping is a spin current generation phenomenon from a magnetic moment into itinerant conduction electron. 
Generally, spin current is a tensor that has current direction and polarization direction. In one dimensional system, the current direction is in $+x$ direction. The polarization direction $\textbf{j}$ of
the spin current can be determined from the following continuity equation for conduction spin.
\begin{align}
\frac{\partial \textbf{j}(x,t)}{\partial x}  +\frac{\partial \textbf{s}(x,t)}{\partial t} = \gamma J\textbf{s}(x,t) \times \textbf{S}(t)\delta(x). \label{Eq.continuity}
\end{align}
We can solve Eq.~\ref{Eq.continuity} by substituting Eq.~\ref{Eq.sigmaReIm} and taking integration over $x$. 
\begin{align}
\textbf{j}(x,t)=& \textbf{j}(-\infty,t)+ J^2{\textbf{S}}\times \dot{\textbf{S}} \int_{-\infty}^x dx \varphi (x)\delta (x) \notag\\
&- J\dot{\textbf{S}}\int_{-\infty}^x dx\chi(x) +J\ddot{\textbf{S}}\int_{-\infty}^x dx\varphi(x).\label{Eq.SpinPump}
\end{align}

\begin{figure}[h]
\centering
\includegraphics[width=\columnwidth]{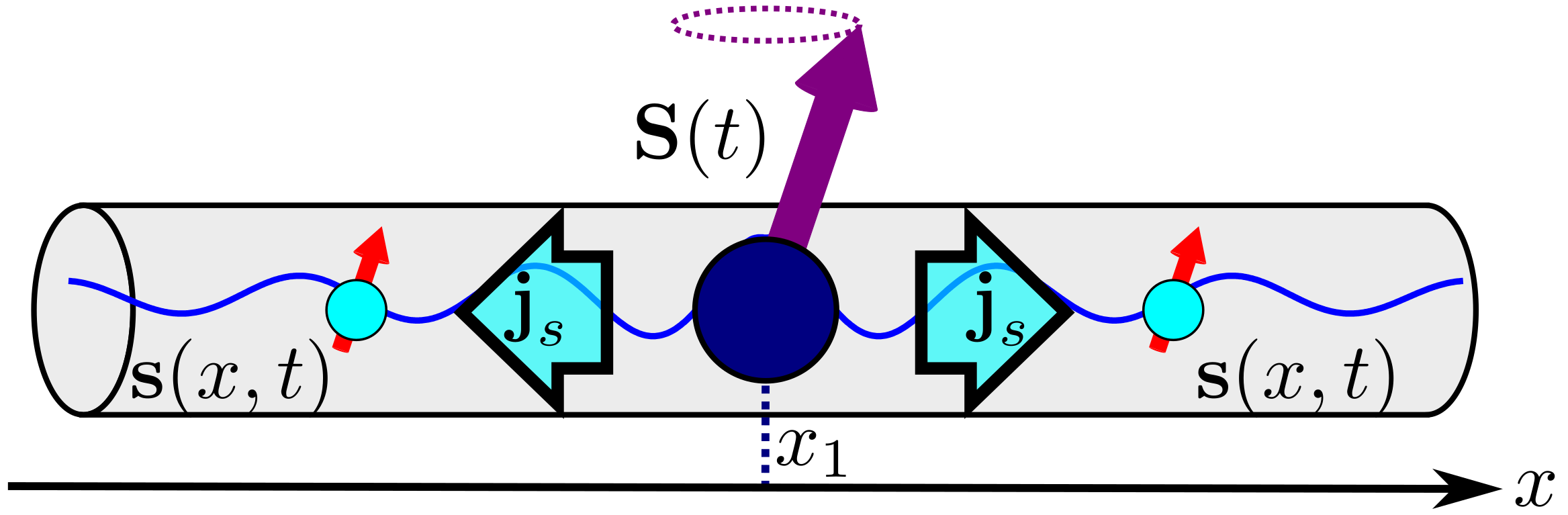}
\caption{Spin pumping from a single impurity $\textbf{S}$ into surrounding conduction spin. 
\label{figspinpumping}}
\end{figure}

The spin current experiment is often observed by ferromagnetic resonance spectrum when external magnetic field $\textbf{B}_\mathrm{ext}=(B_0\cos\omega t, \ B_0\sin\omega t, \ B_z)$ is applied. Here, $B_0\ll B_z$.
The magnitude of the spin current due to $\textbf{B}_\mathrm{ext}$ can be calculated by solving Larmor precession equation of $\textbf{S}$.
\begin{align}
\dot{\textbf{S}}(t)=& -\gamma \textbf{S}(t) \times \left(\textbf{B}_\mathrm{ext}(t) -J \textbf{s}(0,t)\right)\notag\\
=& -\gamma \textbf{S}(t) \times \textbf{B}_\mathrm{ext}(t) -\gamma J\varphi(0) \textbf{S}(t)\times\dot{\textbf{S}}(t). \label{Eq.LLG1impurity}
\end{align}
The last term indicates correspond to the angular momentum loss due to spin pumping. Eq.~\ref{Eq.LLG1impurity} can be linearized by assuming that $S_x,S_y\ll S_z$ as follows.
\begin{equation}
\left[
\begin{array}
[c]{cc}%
\frac{d}{dt} & \omega_0-\alpha_0\frac{d}{dt}\\
-\omega_0+\alpha_0\frac{d}{dt} & \frac{d}{dt}%
\end{array}
\right]  \left[
\begin{array}
[c]{c}%
S_{x}\\
S_{y}
\end{array}
\right]
= \left[
\begin{array}
[c]{c}%
\gamma S_{z}B_y\\
-\gamma S_{z}B_x
\end{array}
\right], \label{Eq.linearSB}
\end{equation}
where $\omega_0=\gamma B_z$ is the resonance frequency and $\alpha_0=\gamma J\varphi(0) S_z$ is damping that arises from adiabatic correction to susceptibility. The solution of Eq.~\ref{Eq.linearSB} is
\begin{equation}
\left[
\begin{array}
[c]{c}
S_{x}\\
S_{y}
\end{array}
\right]
= \frac{\gamma S_zB_0}{\sqrt{(\omega-\omega_0)^2+\alpha_0^2\omega^2}} \left[
\begin{array}
[c]{c}%
\cos \left(\omega t + \tan^{-1} \frac{\alpha_0\omega}{\omega_0-\omega}\right)\\
\sin \left(\omega t + \tan^{-1} \frac{\alpha_0\omega}{\omega_0-\omega}\right)
\end{array}
\right]. \label{Eq.solutionS}
\end{equation}
By using Eq.~\ref{Eq.solutionS}, we can calculate the time-averaged of $\textbf{j}$. Since only $\sin^2$ and $\cos^2$ terms remain, the time-averaged current $\left<\textbf{j}(x,t)\right>_t$ only arise from $\textbf{S}\times\dot{\textbf{S}}$ term.
\begin{align}
\left<\textbf{j}(x),t\right>_t=\left\{
\begin{array}{cc}
\textbf{j}_s(\omega) , & x>0,\\
-\textbf{j}_s(\omega) , & x<0,
\end{array}
\right.\label{Eq.Js}
\end{align}
where 
\begin{equation}
\textbf{j}_s(\omega)=\pm \left<\textbf{j}(\pm\infty,t)\right>_t=\hat{\textbf{z}} \frac{\omega \left(\gamma S_zB_0J\varphi(0)\right)^2}{(\omega-\omega_0)^2+\alpha_0^2\omega^2}.
\end{equation}
Eq.~\ref{Eq.Js} indicates that for $x>0$ spin current is pumped away from the impurity with polarization $\textbf{j}_s(\omega)$. Symmetry of the system dictates that $-\textbf{j}_s(\omega)$ polarization with $+x$ current direction is the same as $+\textbf{j}_s(\omega)$ polarization with $-x$ current direction. Therefore, spin current in $x<0$ is also pumped away from the impurity with polarization $\textbf{j}_s(\omega)$, as illustrated in Fig.~\ref{figspinpumping}.

\subsection{Indirect damping between two impurities}
\label{Sec.indirectdamping}
In Sec.~\ref{Sec.spinpumping}, we show that $\varphi(0)$ term emerges as a damping in the dynamic of a single impurity. 
To take a look at the effect of $\varphi(x)$ on the dynamic of the two impurities, let's take a look at the Larmor precession of $j$-th impurity when we apply an external magnetic field $B_\mathrm{ext}$. 
\begin{align}
\dot{\textbf{S}}_j(t)=-\gamma \textbf{S}_j(t)\times \left(\textbf{B}_\mathrm{ext}(t)-J\textbf{s}(x_j,t)\right), \label{Eq.torque}
\end{align}
Here, $J\textbf{s}$ term arises from the exchange interaction in Eq.~\ref{Eq.Hamiltonian} and $\gamma$ is the gyromagnetic ratio of the impurity spins. For simplicity, we assume that the impurity spins have the same $\gamma$ value.

By, substituting Eq.~\ref{Eq.sigmaReIm} to Eq.~\ref{Eq.torque}, we arrive at the following set of equations
\begin{align}
\dot{\textbf{S}}_j=& -\gamma J\textbf{S}_j\times \textbf{B}_\mathrm{ext} +\gamma J^2\chi(x_j-x_i)  \textbf{S}_j\times \textbf{S}_{i\neq j}(x_j,t)\notag\\ 
&-\gamma J^2 \textbf{S}_j\times \sum_i\dot{\textbf{S}}_i\varphi(x_i-x_j) ,
\end{align}
where $i,j\in\{1,2\}$.
One can see that the sign of $\chi(x_1-x_2)$ determines whether the coupling is ferromagnetic or antiferromagnetic, as expected from the RKKY interaction.
Because of this coupling, the dynamic of the coupled system can be written in the following Landau-Lifshitz-Gilbert equation
\begin{align}
\dot{\textbf{M}}=-\gamma\textbf{M}\times\textbf{B}_\mathrm{ext} + \alpha\textbf{M}\times \dot{\textbf{M}}
\end{align}
where $\textbf{M}=\gamma\left(\textbf{S}_1+\textbf{S}_2\right)$ and 
\begin{align}
\alpha=& J^2\frac{\left(S_1^2+S_2^2\right)\varphi(0)+2S_1S_2\varphi(|x_1-x_2|)}{(S_1+S_2)^2}\notag\\
=& J^2\varphi(0)\frac{\left(S_1^2+S_2^2\right)+2S_1S_2\cos^2\left(k_F|x_1-x_2|\right)}{(S_1+S_2)^2}, \label{Eq.damping}
\end{align}
is the Gilbert damping that arises from the adiabatic correction to susceptibility. Therefore, $\varphi$ is a dissipative term. 
We can see that $\varphi$ induces magnetic damping that fluctuates between the following values
\begin{align}
J^2\varphi(0)\frac{S_1^2+S_2^2}{(S_1+S_2)^2} \leq\alpha\leq J^2\varphi(0).
\end{align}

\section{Conclusion}
\label{Sec:conclusion}
In summary, we calculate the range function for RKKY interaction in one dimension for dynamic magnetic impurities. By considering finite frequency in the dynamic magnetic susceptibility, we show that strong singularity at the $k=q=0$ pole is decomposed to several simple poles, as illustrated in Fig.~\ref{figContour}b. We show that the poles can be used to evaluate the dynamic susceptibility for small $\omega\ll k_F^2$.

In the adiabatic limit, the leading term of correction due to frequency dependency gives an imaginary term $\chi(x,\omega)= \chi(x)+i\omega \varphi(x)$. $\varphi(x)$ also has $2k_Fx$ oscillation but there is no change in sign, as illustrated in Fig.~\ref{figReIm}. This term induces a correction in spin polarization of the conduction electron that mediates the RKKY interaction. The additional term is dissipative and therefore generates magnetic damping $\alpha$ that depends on the distance between two impurities (see Eq.~\ref{Eq.damping}). In the case of single impurity, the angular momentum loss due to the damping torque is transferred into the conduction spin as spin current.

The dependency of $\alpha$ to distance and Fermi momentum can be utilized to manipulate the spin impurity. Further study on the dynamic susceptibility may shed light on the spin transfer mechanism and other spin related effects in one dimensional system.


\section*{Declaration of competing interest} 
The author declare that they have no known competing financial
interests or personal relationships that could have appeared to
influence the work reported in this paper.

\appendix

\section{Alternative derivation of the imaginary part of the dynamic susceptibility.} \label{Sec:appendix-spatial-ImSusc}
Substituting Sokhotski-Plemelj formula (Eq.~\ref{Eq.PS}) to Eq.~\ref{Eq.chiAdiabatic}, we arrive at the following expression for imaginary susceptibility.
\begin{align*}
\mathrm{Im}\chi( x,\omega)=&\int \frac{dq}{2\pi}e^{i q\cdot x}\mathrm{Im}\chi( q,\omega)\\
\mathrm{Im}\chi( q,\omega)=&-\pi\int \frac{dk}{2\pi} \left(f_k-f_{k+ q}\right) \notag\\
&\times\delta\left(\epsilon_{k+ q}-\epsilon_k+\omega+i\eta\right). 
\end{align*}
By shifting  $f_{k+q}$ to $f_k$, we can show that in adiabatic regime, it is linearly proportional to the frequency 
\begin{align*}
\mathrm{Im}\chi(q,\omega)=&-\pi\int \frac{dk}{2\pi} f_ k\Bigg(\delta\left(\epsilon_{ k+ q}-\epsilon_ k+\omega\right)\\
&-\delta\left(\epsilon_{ k+ q}-\epsilon_ k-\omega\right)\Bigg)\\
\lim_{\omega\to 0} \mathrm{Im}\chi(q,\omega)=& -\omega \int dk f_ k\delta'\left(\epsilon_{ k+ q}-\epsilon_ k\right),
\end{align*}
where $\delta'(x)$ is the first derivation of $\delta$ function. The small $\omega\ll k_F^2$ limit can be used to evaluate the integral.
\begin{align*}
&\mathrm{\operatorname{Im}}\chi(q,\omega) 
=\int_{-k_F}^{k_F} dk\left[\delta\left(2kq -q^2+\omega\right) - \delta\left(2kq +q^2+\omega\right)\right]\\
&=\int_{-k_F}^{k_F} \frac{dk}{2|q|} \left[\delta\left(k -\frac{q^2-\omega}{2q^2}\right) -\delta\left(k +\frac{q^2+\omega}{2q^2}\right)\right]\\
&=\frac{1}{2|q|}   \left[\Theta\left(\left|k_F - \frac{q^2-\omega}{2q^2} \right| \right) -\Theta \left(\left|k_F  - \frac{q^2+\omega}{2q^2} \right| \right) \right].
\end{align*}
The first step function is 1 when
\begin{align*}
-k_F-\sqrt{k_F^2+\omega}<q<k_F-\sqrt{k_F^2+\omega}\\
-k_F+\sqrt{k_F^2+\omega}<q<k_F+\sqrt{k_F^2+\omega},
\end{align*}
otherwise it is zero. On the other hand, the second step function is 1 when
\begin{align*}
-k_F-\sqrt{k_F^2-\omega}<q<-k_F+\sqrt{k_F^2-\omega}\\
k_F-\sqrt{k_F^2-\omega}<q<k_F+\sqrt{k_F^2-\omega},
\end{align*} 
otherwise it is zero.

By reordering the inequalities, we can conclude that the imaginary susceptibility is non zero on the following ranges: 
\begin{align*}
-k_F-\sqrt{k_F^2+\omega}<q<-k_F-\sqrt{k_F^2-\omega} \\
-k_F+\sqrt{k_F^2-\omega}<q<k_F-\sqrt{k_F^2+\omega}\\
-k_F+\sqrt{k_F^2+\omega}<q<k_F-\sqrt{k_F^2-\omega}\\
k_F+\sqrt{k_F^2-\omega}<q<k_F+\sqrt{k_F^2+\omega}.
\end{align*}

By evaluating its inverse Fourier transform, we arrive at the same expression as Eq.~\ref{Eq.varphi}.
\begin{align*}
&\lim_{\omega\ll \epsilon_F}\mathrm{\operatorname{Im}}\chi(x,\omega)=\lim_{\omega\ll \epsilon_F}\int_0^{\infty} \frac{dq \cos qx}{ 2\pi}  \mathrm{\operatorname{Im}}\chi(q,\omega)\\
&=  \left( \int_{\frac{\omega}{2 k_F}-\frac{\omega^2}{8 k_F^3}}^{\frac{\omega}{ 2k_F}+\frac{\omega^2}{8 k_F^3}} + \int_{2k_F-\omega/ 2k_F}^{2k_F+\omega/ 2k_F} \right)\frac{dq \cos qx}{ 2\pi q} \\
&= \lim_{\omega\to 0}  \left( \frac{\omega^2}{4 k_F^3} \frac{\cos \frac{\omega}{ k_F}}{\pi{\omega}/{k_F}} + \frac{\omega}{ k_F}\frac{ \cos 2k_Fx}{ 4\pi k_F} \right) \\
&= \frac{ \omega}{4\pi k_F^2} \left(1+\cos 2k_Fx \right)= \frac{ \omega}{2\pi k_F^2} \cos^2 k_Fx
\end{align*}


\end{document}